\documentclass[aps,nofootinbib]{revtex4-2}

\usepackage{hyperref}
\usepackage{braket}
\usepackage{amssymb}
\usepackage{amsmath}
\usepackage{amsthm}
\usepackage{amscd}
\usepackage[all,arc,knot,matrix]{xy}

\usepackage{graphicx}
\usepackage{mathpazo}
\usepackage{CJK}

\theoremstyle{definition}

\theoremstyle{remark}

\newcommand{\abs}[1]{\lvert#1\rvert}

\begin{document}
\begin{CJK*}{UTF8}{bsmi}
\title{Degenerate Limits of Scalar-Tensor Gravity}
\author{Bekir Bayta\c{s}}
\email{bekirbaytas@iyte.edu.tr}
\affiliation{Center for Relativity and Gravitation,
School of Physics and Astronomy,
Beijing Normal University, Beijing 100875, China}
\affiliation{Department of Mathematics,
\.{I}zmir Institute of Technology,
G\"ulbah\c{c}e, Urla 35430, \.{I}zmir, Turkey}

\author{Xiao-Kan Guo}
\email{kankuohsiao@whu.edu.cn}
\affiliation{Center for Relativity and Gravitation,
School of Physics and Astronomy,
Beijing Normal University, Beijing 100875, China}
\affiliation{School of Mathematics and Physics,
Yancheng Institute of Technology,
Jiangsu 224051, China}
\date{\today}

\begin{abstract}
We study the degenerate limits of the scalar-tensor theory of gravity in its Hamiltonian connection dynamics formulation. The degenerate limit of the spatial triad is effectuated by a local scaling transformation of the triad, which leads to the degenerate limits of all other geometric quantities and relations in the connection dynamics. We derive the constraints for the degenerate scalar-tensor theory of gravity and discuss their physical implications.
\end{abstract}

\maketitle
\section{Introduction}
The introduction of the connection variables by Ashtekar {\it et al.} \cite{Sen82,Ash86,Bar95,Imm97} for general relativity (GR) has transformed  the Hamiltonian approach to gravity into a theory of connection dynamics. A quantization of GR based on connection dynamics has led to the rapid development of loop quantum gravity (LQG) and spin foam models\cite{Rov04,Thi07,RV15}. Both the connection dynamics and the loop quantization can, in principle, be generalized to many other modified theories of gravity, particularly the metric-affine generalizations of GR, where metric variables can be transformed into connection variables. Indeed, the loop quantization of scalar-tensor gravity, a typical modified theory, has been successfully implemented in~\cite{ZM11a,ZM11b,ZM11c,ZM12}, showcasing the viability of both the connection dynamics formulation and LQG quantization techniques in this context.

In the early studies,  an interesting degenerate phase of the connection dynamics based on the complex Ashtekar variables is identified as an extension of GR. This degeneracy is set by taking the determinant of the spatial triad to zero, which results in the non-invertibility of the triad, and hence the connection dynamics formulation cannot be transformed back to the ADM formulation. Therefore, only the nondegenerate phase of the connection dynamics (with the reality condition satisfied) is argued to be  equivalent to GR. There are actually many such  degenerate extensions of GR \cite{JR92}. 
Later investigations (cf. \cite{YSN97,BJ97,ML98}), however, reveal many intimate relations between the degenerate and the nondegenerate phases of connection dynamics. 

Since such a degeneracy can be formally performed within a connection dynamics, one naturally asks whether there is a way of realizing a degenerate phase within GR, given that the Holst action in terms of the real Barbero-Immirzi  variables differs from the Palatini action only by a  term vanishing on half-shell \cite{Hol96}. 
In a recent work by Sengupta \cite{Sen25}, a new degenerate limit of the connection dynamics of the Palatini action for GR is taken, and it is found that the resulting Hamiltonian theory recovers the Hamiltonian formulation of Carroll gravity in connection formulation \cite{Sen23}. This result explicitly realizes the degenerate limit within GR, as the Carroll gravity is the ultrarelativistic ($c\rightarrow0$) limit of GR \cite{BGRRt17}. Importantly, this result also indicates that one cannot directly start with some limits of the ADM formulation of GR and try to get the corresponding connection dynamics or LQG for this type of gravity, since the degenerate triad is no longer invertible. Notice that the Carrollian limit of the Holst action has been considered in the recent work \cite{BMVV25}, but it is only considered for the Holst action written in terms of  spin connections and the main focus is on the Carrollian limit of the Holst topological term which results in the action of the Husain-Kuchar model.

Inspired by the successful generalization of connection dynamics from GR to scalar-tensor theory of gravity,
 we study in this note the degenerate limits for the canonical scalar-tensor theory of gravity by using also the scaling approach as in \cite{Sen25}. The previous degenerate limit  only refers to the degenerate  triad. We would like to see how other geometric quantities or relations, such as the constraints, as well as the scalar field contributions change in the degenerate limit. By finding how exactly the geometric quantities scale with the degenerate limit, we are able to derive the constraints for the degenerate scalar-tensor gravity.\footnote{Here, the degenerate scalar-tensor gravity is {\it not} the same as the degenerate higher-order scalar-tensor theory (DHOST) where the degeneracy condition is imposed on the kinetic matrix to avoid the Ostrogradski instability.} 
For the geometric part of the constraints, we confirm some similar structures as those in the degenerate GR \cite{Sen25}, but we also found significant differences.

We begin in Section \ref{S2} with a brief review of the connection dynamics of scalar-tensor theory. In Section \ref{S3} we analyze the degenerate limits of scalar-tensor theory by considering two sets of local scaling transformations of the connection dynamics. An explicit example of the degenerate limits of  Bianchi I model in Brans-Dicke gravity is considered in Section \ref{S4}. The final Section \ref{S5} concludes.

\section{Connection dynamics of scalar-tensor gravity}\label{S2}
In this  section, we review the connection dynamics of the scalar tensor theory of gravity. 

In a general scalar-tensor theory of gravity, a scalar field $\phi$ is non-minimally coupled to the metric $g_{ab}$ of the spacetime $M$.
The general action for the scalar-tensor theory of gravity in the Jordan frame reads 
\begin{equation}\label{1}
S[g,\phi]=\int_Md^4x\sqrt{-g}\Bigl[\frac{1}{2}\Bigl(\phi R-\frac{w(\phi)}{\phi}\partial_a\phi\partial^a\phi\Bigr)-V(\phi)\Bigr],
\end{equation}
where $\omega(\phi)$ is a coupling parameter depending on $\phi$ and $V(\phi)$ is an arbitrary potential function for $\phi$. There are several notable special cases of the scalar-tensor theory of gravity \cite{SF10}: The case in which $w=0$ in \eqref{1} is the metric $f(R)$ gravity, while the case with  $w=-\frac{3}{2}$ corresponds to the Palatini $f(R)$ gravity. Another special case is the Brans-Dicke gravity with $w=\text{const.}$ and $V=0$.

The Hamiltonian analysis of the action \eqref{1} in the ADM formalism can be found in \cite{ZM11c}, where a particular value of the coupling parameter, $w=-\frac{3}{2}$ is singled out as a special case in which there is an extra constraint. The connection dynamics for the scalar-tensor theories in both sectors (i.e. $w\neq-\frac{3}{2}$ and $w=-\frac{3}{2}$) can be obtained the from the ADM formulation through canonical transformations \cite{ZM11c}. An equivalent approach to the connection dynamics of scalar-tensor theories in the Jordan frame is to start with the following first-order action \cite{ZGHM13}
\begin{align}
S[e,\omega,\phi]=\int_Md^4x\frac{1}{2}\Bigl[&\phi ee^a_Ie^b_JR_{ab}^{\phantom{ab}IJ}-2ee^a_Ie^b_J\omega_a^{IJ}\partial_b\phi+ee_I^{[a}e_J^{b]}\partial_a(e_b^Ie^{cJ}\partial_c\phi)+\nonumber\\
&+\frac{w(\phi)}{\phi}e\partial_a\phi\partial^a\phi-2eV(\phi)+ee^a_Ie^b_J\frac{1}{\gamma}~^*R_{ab}^{\phantom{ab}IJ}\Bigr]\label{2}
\end{align}
in terms of the tetrad $e^a_I$ and the $SL(2,\mathbb{C})$ spin connection $\omega_a^{IJ}$. In \eqref{2}, the indices $a,b,...$ represent the spacetime indices, and $I,J,...$ are the internal Lorentzian indices. We have also used in \eqref{2} the curvature $R_{ab}^{\phantom{ab}IJ}=\partial_{[a}\omega_{b]}^{IJ}+\omega^{IK}_{[a}\omega_{b]K}^{\phantom{b]K}J}$, its Hodge dual $~^*R_{ab}^{\phantom{ab}IJ}$, and the Barbero-Immirzi parameter $\gamma\in\mathbb{R}$. 

To obtain the connection dynamics, we should consider  the  $(3+1)$-decomposition of spacetime, $M=\Sigma\times\mathbb{R}$. Let $n^a$ be a unit vector normal to $\Sigma$, and let $q_{ab}$ the metric on $\Sigma$, then we have $g_{ab}=q_{ab}-n_an_b$. We  denote the spatial triad  by
\begin{equation}
E^a_I=q^a_be^b_I.
\end{equation}
In the following, as in LQG, we restrict ourselves to the $SU(2)$ sector of $SL(2,\mathbb{C})$ on $\Sigma$, and we use the indices $i,j,...$ as internal $SU(2)$ indices.
Let $E=\sqrt{q}$ be the square root of the determinant of $q_{ab}=E_a^iE_{bi}$ , then the densitized triad is
\begin{equation}\label{4}
\tilde{E}^a_i=EE^a_i.
\end{equation}
 Since there exists a non-minimal coupling between the scalar field and geometry, we can introduce a new form of the extrinsic curvature
\begin{equation}\label{5}
K^i_a=\phi\omega_a^{i0}+\frac{1}{2}E^i_an^c\partial_c\phi,
\end{equation}
with its spatial component denoted as $\tilde{K}^i_a=q^b_aK^i_b$ . We also define the Barbero-Immirzi connection
\begin{equation}\label{66}
A^i_a=\Gamma^i_a+\gamma \tilde{K}^i_a
\end{equation}
with $\Gamma^i_a=-\frac{1}{2}\epsilon^i_{\phantom{i}jk}\omega^{jk}_a$, the $SU(2)$ spin connection on $\Sigma$. The canonical pair $(A^i_a,\tilde{E}^a_i)$ span the phase space for the canonical scalar-tensor theories.

After fixing the gauge such that $n^ae_{aI}=(1,0,0,0)$, we can perform the Hamiltonian analysis of the first-order action \eqref{2} \cite{ZGHM13}. The resulting Hamiltonian can be expressed as 
\begin{equation}\label{6}
H=\int d^3x\bigl(\Lambda^i\mathcal{G}_i+N^a\mathcal{C}_a+\underline{N}\mathcal{C}\bigr).
\end{equation}
The $\Lambda^i,N^a,\underline{N}$ in \eqref{6} are Lagrange multipliers, and in particular, $N^a$ is the shift vector and $\underline{N}=N/E$ is the densitized lapse.  The other terms $\mathcal{G}_i,\mathcal{C}_a,\mathcal{C}$ in \eqref{6} are respectively the Gaussian, vector, and scalar constraints. When $w\neq-\frac{3}{2}$, these constraints are expressed as
\begin{align}
\mathcal{G}_i&=\partial_b\tilde{E}^b_j+\epsilon_{jl}^{\phantom{jl}m}A^l_b\tilde{E}^b_m,\label{8}\\
\mathcal{C}_a&=\tilde{E}^b_j\nabla_{[a}\tilde{K}^j_{b]}+\pi\partial_a\phi,\label{9}\\
\mathcal{C}&=\frac{\phi}{2}\tilde{E}^a_i\tilde{E}^b_j\epsilon^{ij}_{\phantom{ij}k}\Bigl(R_{ab}^{\phantom{ab}j}-\frac{1}{\phi^2}\epsilon^k_{\phantom{k}lm}\tilde{K}^l_a\tilde{K}^m_b\Bigr)+\tilde{E}^a_i\tilde{E}^{bi}\nabla_a\nabla_b\phi+\frac{1}{2}\frac{w}{\phi}\tilde{E}^a_i\tilde{E}^{bi}\partial_a\phi\partial_b\phi+\nonumber\\
&~~~~+\frac{\phi}{2(w+\frac{3}{2})}\Bigl(\pi+\frac{1}{\phi}\tilde{E}^b_j\tilde{K}^j_b\Bigr)^2+E^2V(\phi),\label{10}
\end{align}
where $\nabla_a$ is the covariant derivative on $\Sigma$ such that $\nabla_aE^a_i=0$, $R_{ab}^{\phantom{ab}j}$ is the curvature calculated from $\Gamma^i_a$, and $\pi$ is the conjugate momentum of the scalar field $\phi$:
\begin{equation}\label{11}
\pi=-\frac{1}{\phi}\tilde{E}^b_j\tilde{K}^j_b+\frac{(w+\frac{3}{2})}{\phi\underline{N}}(\partial_0\phi-N^a\partial_a\phi).
\end{equation}
When $w=-\frac{3}{2}$, there is an extra conformal constraint (from \eqref{10} and \eqref{11}),
\begin{equation}
S=\pi\phi+\tilde{E}^b_j\tilde{K}^j_b.
\end{equation}
So in this case, the Hamiltonian becomes
\begin{equation}
H=\int d^3x\bigl(\Lambda^i\mathcal{G}_i+N^a\mathcal{C}_a+\underline{N}\mathcal{C}_0+\lambda S\bigr),
\end{equation}
with a simplified scalar constraint
\begin{equation}\label{14}
\mathcal{C}_0=\frac{\phi}{2}\tilde{E}^a_i\tilde{E}^b_j\epsilon^{ij}_{\phantom{ij}k}\Bigl(R_{ab}^{\phantom{ab}j}-\frac{1}{\phi^2}\epsilon^k_{\phantom{k}lm}\tilde{K}^l_a\tilde{K}^m_b\Bigr)+\tilde{E}^a_i\tilde{E}^{bi}\nabla_a\nabla_b\phi-\frac{3}{4}\frac{1}{\phi}\tilde{E}^a_i\tilde{E}^{bi}\partial_a\phi\partial_b\phi+E^2V(\phi).
\end{equation}

Note that the above formulation of the connection dynamics for scalar-tensor theories is presented within the Jordan frame. But this can be transformed to a formulation in the Einstein frame through a canonical transformation (cf. \cite{ZGHM13,CM09,BBM21}).
\section{Degenerate limits of scalar-tensor theory}\label{S3}
We turn to the degenerate limits of scalar-tensor theories in the connection dynamics formulation.
Here, by degenerate limit we mean the limit that (the matrix of) the spatial triad is degenerate, i.e. with vanishing determinant. To get control over such a limit, let us consider, as in \cite{Sen25}, a scaling transformation on the triad
\begin{equation}\label{1515}
E^i_a \mapsto \delta_{(a)} {E'}^i_a,
\end{equation}
where  $\delta_{(a)}$ is a  scaling parameter. The index $(a)$ means that it is not summed over when this  index is repeated. 
But, because the scaling parameter has the index $(a)$, the scalings are anisotropic. In other words, this local scaling parameter $\delta_{(a)}$ only depends on the direction, instead of depending on the local coordinates; consequently, the spatial derivatives of $\delta_{(a)}$ should vanish, e.g.  $\partial_a\delta_{(a)}=0$.
 We can define 
\begin{equation}
\delta=\prod_a\delta_{(a)}, 
\end{equation}
so that 
the degenerate limit is the limit that $\delta\rightarrow0$, since $E=\sqrt{q}=\delta\sqrt{q'}=\delta E'$. Note that an individual $\delta_{(a)}$ does not need to vanish in this limit.
Under such a scaling transformation, the densitized triad \eqref{4} transforms  as
\begin{equation}\label{17}
\tilde{E}^a_i \mapsto \frac{\delta}{\delta_{(a)}}\tilde{E'}^a_i,
\end{equation}
where the $\delta_{(a)}$ on the denominator comes from the inversion of $E^i_a$.  Recall that the Carrollian limit can be realized by the limit $\epsilon\rightarrow0$ in the scaling transformations on the triads as $e^0_a=\epsilon {e'}^0_a$ and $e^a_0=\frac{1}{\epsilon} {e'}^a_0$ \cite{BGRRt17}. Although these scaling transformations and \eqref{1515} are different in the internal indices, the overall effect is the same, i.e. the determinant of the tetrad tends to $0$ in both limits. 

The degenerate limits of the triads and densitized triads also change other geometric quantities such as connections. However, the way in which other quantities change under the degenerate limit is not unique. In \cite{Sen25} two sets of  degenerate limits are considered, and it is found that the first set of degenerate limits changes GR to the electric Carroll gravity, while the second set changes GR into the magnetic Carroll gravity. In the following we consider these two sets of limits for scalar-tensor gravity.
\subsection{The first limit}
We are interested in the degenerate limits of the canonical scalar-tensor theory of gravity. To this end, we first look at the scaling transformations for various quantities appeared in the constraints \eqref{8}\eqref{9}\eqref{10}. First, if we want to  keep the symplectic structure (or the Poisson brackets) of the phase space spanned by $(A^i_a,\tilde{E}^a_i)$,  the Barbero-Immirzi connection should scale as
\begin{equation}\label{18}
A^i_a \mapsto \frac{\delta_{(a)}}{\delta}{A'}^i_a.
\end{equation}
 With these transformations in hand, we see that  \eqref{8} becomes
\begin{equation}
\mathcal{G}_i=\frac{\delta}{\delta_{(a)}}\partial_b\tilde{E'}^b_j+\epsilon_{jl}^{\phantom{jl}m}{A'}^l_b\tilde{E'}^b_m.
\end{equation}
In the degenerate limit $\delta\rightarrow0$, this tends to
\begin{equation}\label{20}
\mathcal{G'}_i=\epsilon_{jl}^{\phantom{jl}m}{A'}^l_b\tilde{E'}^b_m.
\end{equation}
On the other hand, due to the the definition \eqref{66}, the scaling transformations of the extrinsic curvature $\tilde{K}^i_a$ and the spatial $SU(2)$ connections $\omega^{ij}_a$ are
 obviously the same as that of $A^i_a$, as in  \eqref{18},
\begin{equation}
\tilde{K}^i_a=\frac{\delta_{(a)}}{\delta}{\tilde{K'}}^i_a,\quad \omega^{ij}_a=\frac{\delta_{(a)}}{\delta}{\omega'}^{ij}_a.
\end{equation}
Let us assume that the scalar field $\phi$ does not change under the scaling transformation. 
If we want to keep the symplectic structure for $(\phi,\pi)$,  the conjugate momentum $\pi$ given by \eqref{11}  should not change under the scaling transformation, which requires that $N=\delta N'$ and $N^a={N'}^a$. However, since there is a covariant derivative $\nabla_a$ in \eqref{9}, there will be a scaling factor $\frac{\delta(a)}{\delta}$ in front of the connection (in the covariant derivative), so the scaling transformation $N^a={N'}^a$ could not absorb this factor, which would result in a divergent behaviour inside the degenerate limit of the covariant derivative. To avoid the possible divergence, we still assume that, as in \cite{Sen25},
\begin{equation}\label{2222}
N=\delta N', \quad N^a=\frac{\delta}{\delta_{(a)}}{N'}^a.
\end{equation}
 Then, under the above assumptions, the vector constraint only contains a term from the covariant derivative $\nabla_a$:
 \begin{equation}\label{22}
 \mathcal{C}'_a=\tilde{E'}^b_j{\omega'}_{i[a}^{j}\tilde{K'}^i_{b]}
 \end{equation}
 where a factor $\frac{\delta(a)}{\delta}$ has been cancelled by $N_a=\frac{\delta}{\delta_{(a)}}{N'}^a$ (so that the terms involving Levi-Civita connections also tend to $0$ in the degenerate limit).

  Furthermore, comparing the curvatures  $R_{ab}^{\phantom{ab}j}$ and ${R'}_{ab}^{\phantom{ab}j}=\partial_{[a}{\omega'}_{b]}^{ij}+{\omega'}^{ik}_{[a}{\omega'}_{b]k}^{\phantom{b]k}j}$, we see that it scales as $R_{ab}^{\phantom{ab}j}\sim\frac{1}{\delta}{R'}_{ab}^{\phantom{ab}j}$, where the symbol $\sim$ means that there are some other $\delta_{(\cdot)}$. Then, the term $\tilde{E}^a_i\tilde{E}^b_j\epsilon^{ij}_{\phantom{ij}k}R_{ab}^{\phantom{ab}j}$ in \eqref{10} scales as $\sim\delta$, so that this term tends to $0$ in the degenerate limit. As a consequence, the scalar constraint \eqref{10} in this limit becomes
\begin{align}
\mathcal{C}'&=-\frac{1}{2\phi}\epsilon^{ij}_{\phantom{ij}k}\epsilon^k_{\phantom{k}lm}\tilde{E'}^a_i\tilde{E'}^b_j\tilde{K'}^l_a\tilde{K'}^m_b+\frac{\phi}{2(w+\frac{3}{2})}\Bigl(\pi'+\frac{1}{\phi}\tilde{E'}^b_j\tilde{K'}^j_b\Bigr)^2=\nonumber\\
&=-\frac{1}{2\phi}\epsilon^{ij}_{\phantom{ij}k}\epsilon^k_{\phantom{k}lm}\tilde{E'}^a_i\tilde{E'}^b_j\tilde{K'}^l_a\tilde{K'}^m_b+\frac{w+\frac{3}{2}}{2\phi\underline{N'}^2}(\partial_0\phi)^2.
\label{23}
\end{align}
Notice that the Lagrange multiplier for the scalar constraint is $\underline{N}=N/E$, so we do not need to absorb a factor $1/\delta$ in to $N$ as in \cite{Sen25}.
In the same limit, the Hamiltonian constraint \eqref{14} for the case with $w=-\frac{3}{2}$ becomes
\begin{equation}\label{24}
\mathcal{C}'_0=-\frac{1}{2\phi}\epsilon^{ij}_{\phantom{ij}k}\epsilon^k_{\phantom{k}lm}\tilde{E'}^a_i\tilde{E'}^b_j\tilde{K'}^l_a\tilde{K'}^m_b.
\end{equation}
Thus, in the degenerate limit, the constraints \eqref{8}\eqref{9}\eqref{10} tend to \eqref{20}\eqref{22}\eqref{23}(or \eqref{24}),
\begin{align}
 \mathcal{G'}_i&=\epsilon_{jl}^{\phantom{jl}m}{A'}^l_b\tilde{E'}^b_m,\\
\mathcal{C}'_a&=\tilde{E'}^b_j{\omega'}_{i[a}^{j}\tilde{K'}^i_{b]},\\
\mathcal{C}'&=-\frac{1}{2\phi}\epsilon^{ij}_{\phantom{ij}k}\epsilon^k_{\phantom{k}lm}\tilde{E'}^a_i\tilde{E'}^b_j\tilde{K'}^l_a\tilde{K'}^m_b+\frac{w+\frac{3}{2}}{2\phi\underline{N'}^2}(\partial_0\phi)^2.
\end{align}
We  observe that the gravitational part of these constraints retain the same form of those for the electric Carroll gravity (cf. \cite{Sen25,Sen23}).

The Hamiltonian constraint in the form of \eqref{23} allows deparametrization of the scalar field $\phi$. Indeed, the second term in \eqref{23}, together with the relation \eqref{11}, can be transformed in  the degenerate limit  to $\frac{\phi}{2(w+\frac{3}{2})}\Bigl(\pi+\frac{1}{\phi}\tilde{E}^b_j\tilde{K}^j_b\Bigr)^2$. In other words, we can solve the constraint $\mathcal{C}'=0$ for $\pi$, and we can write $\mathcal{C}'=\pi-C$, the deparametrized form. The deparametrization of the scalar field is important for the relational quantization, cf. \cite{YLLM25}.

Another simple consequence of the degenerate limit is that it only sends various terms from the connection dynamics of scalar-tensor theory to zero, but not creates new terms, so the constraint algebra remains closed at the level of first class.

\subsection{The second limit}
Next, let us consider an alternative degenerate limit advocated in \cite{Sen25} with a different scaling  transformation for $A_a^i$:
\begin{equation}\label{25}
A^i_a={\delta_{(a)}}{A'}^i_a, \quad\tilde{E}^a_i=\frac{\delta}{\delta_{(a)}}\tilde{E'}^a_i.
\end{equation}
Obviously, these scaling transformations do not preserve the symplectic structure of the phase space spanned by $(A^i_a,\tilde{E}^a_i)$. As before,  the scaling transformations of $\tilde{K}^i_a$ and  $\omega^{ij}_a$ are determined by the definition \eqref{66}:
\begin{equation}
\tilde{K}^i_a={\delta_{(a)}}{\tilde{K'}}^i_a,\quad \omega^{ij}_a={\delta_{(a)}}{\omega'}^{ij}_a.
\end{equation}
An additional assumption is that the spatial derivatives transform under the scaling as
\begin{equation}\label{27}
\partial_a=\frac{\delta_{(a)}}{\delta}\partial'_a,
\end{equation}
such that $\tilde{E}^a_i\partial_a$ is invariant under the scaling transformation.\footnote{The transformation \eqref{18} in fact allows a scaling transformation for the spatial derivatives as $\partial_a=\delta_{(a)}\partial'_a$. But this scaling of spatial derivative does not affect the degenerate limit, because all the spatial derivative terms in \eqref{20}\eqref{22}\eqref{23}(or \eqref{24}) tend to zero. From \eqref{25} and \eqref{27}, we see that the alternative degenerate limit puts the scaling factor $\frac{1}{\delta}$ in front of $\partial'_a$ instead of ${A'}^i_a$. }
 This assumed scaling transformation implies that the Gaussian constraint could transform as
\begin{equation}
\mathcal{G''}_i=\frac{\delta}{\delta_{(a)}}\partial_b\tilde{E'}^b_j+\delta\epsilon_{jl}^{\phantom{jl}m}A^{\prime l}_b\tilde{E'}^b_m
=\partial'_b\tilde{E'}^b_j+\delta\epsilon_{jl}^{\phantom{jl}m}A^{\prime l}_b\tilde{E'}^b_m,
\end{equation}
which, in the degenerate limit $\delta\rightarrow0$, becomes
\begin{equation}
\mathcal{G''}_i=\partial'_b\tilde{E'}^b_j.
\end{equation}

As for the vector constraint, 
if we still assume the scaling transformations \eqref{2222} for $N$ and $N^a$, then the vector constraint obviously vanishes in the degenerate limit:
\begin{equation}\label{34}
\mathcal{C''}_a=0.
\end{equation}
This is because, by taking out a common factor $\frac{\delta_{(a)}}{\delta}$ to cancel the factor $\frac{\delta}{\delta_{(a)}}$ from $N^a$, we can see that each term will have a factor $\delta$ which sends every term to $0$ in the degenerate limit. 
 Let us also consider the alternative case in which the shift and lapse functions do not change under the scaling transformation: The covariant derivative term obviously scales as $\sim\delta$, and the conjugate momentum $\pi$ also scales as $\sim\delta$ where the factor $\delta$ comes from the $E$ in $\underline{N}$. In either approaches, we obtain the transformed constraint  \eqref{34}.
 
 In the Hamiltonian constraint, neither the $\omega_a^{ij}$ nor $\tilde{K}^i_a$ could cancel the factor $\delta^2$ from  $\tilde{E}^a_i\tilde{E}^b_j$, so the curvature terms tend to $0$ in the degenerate limit. Due to the assumption \eqref{27}, we have $\tilde{E}^a_i\partial_a=\tilde{E'}^a_i\partial'_a$. Since the covariant derivatives act on scalars as ordinary derivatives, what remain in the Hamiltonian constraint in the degenerate limit for $w\neq-\frac{3}{2}$ are just the scalar-dependent terms:
 \begin{equation}
 \mathcal{C''}=\tilde{E'}^a_i\tilde{E'}^{bi}\partial'_a\partial'_b\phi+\frac{1}{2}\frac{w}{\phi}\tilde{E'}^a_i\tilde{E'}^{bi}\partial'_a\phi\partial'_b\phi+\frac{w+\frac{3}{2}}{2\phi\underline{N'}^2}(\partial_0\phi)^2.
 \end{equation}
 For for $w=-\frac{3}{2}$, this scalar constraint simplifies to
  \begin{equation}
 \mathcal{C''}_0=\tilde{E'}^a_i\tilde{E'}^{bi}\partial'_a\partial'_b\phi-\frac{3}{4}\frac{w}{\phi}\tilde{E'}^a_i\tilde{E'}^{bi}\partial'_a\phi\partial'_b\phi.
 \end{equation}
 
 We collect the constraints for $w\neq-\frac{3}{2}$ in the degenerate limit:
\begin{align}
\mathcal{G''}_i&=\partial'_b\tilde{E'}^b_j,\\
\mathcal{C''}_a&=0,\label{38}\\
\mathcal{C''}&=\tilde{E'}^a_i\tilde{E'}^{bi}\partial'_a\partial'_b\phi+\frac{1}{2}\frac{w}{\phi}\tilde{E'}^a_i\tilde{E'}^{bi}\partial'_a\phi\partial'_b\phi+\frac{w+\frac{3}{2}}{2\phi\underline{N'}^2}(\partial_0\phi)^2.\label{39}
\end{align}
The remaining terms consists of $\tilde{E}^a_i$ and $\partial\phi$. These are conjugate momenta of $A$ and $\phi$ respectively, so there are no nontrivial Poisson brackets between them. Just as in \cite{Sen25}, there would be no ordering problem in the subsequent qunatization. Also, the diffeomorphism constraint \eqref{38} tells us that it is satisfied already at the classical level. But the Hamiltonian constraint no longer share the similar structure with the magnetic Carroll gravity. Due to the last term in the Hamiltonian constraint \eqref{39}, one can still perform a simple deparametrization of the theory by treating the scalar field $\phi$ as time.

As a side remark, the ``extremely'' degenerate limit in which the assumption \eqref{27} is absent would send those terms containing $\delta'_a$ to $0$, such that the remaining non-vanishing constraint is simply the Hamiltonian constraint
\begin{equation}
\mathcal{C''}=\frac{w+\frac{3}{2}}{2\phi\underline{N'}^2}(\partial_0\phi)^2,
\end{equation}
which means only the scalar field survives the limit. It is interesting to compare this case with  the opposite case in which there is no Hamiltonian constraint but has the other two constraints such as the Husain-Kuchar model. In view of the recent  work \cite{BMVV25}, it is possible for the degenerate limit to be linked to the gravitational part of the Carrollian limit and in complementarity with the Husain-Kuchar part.
Note that this  scalar-field dynamical term also vanishes when $w=-\frac{3}{2}$, which is not physically interesting as nothing remains in the limit. 
\section{Degenerate limit of Brans-Dicke Bianchi I cosmology}\label{S4}
As an explicit example of the degenerate limits of scalar-tensor theories, let us consider the Bianchi I cosmological model in Brans-Dicke gravity and its degenerate limit. As mentioned before, Brans-Dicke gravity is a special case of the scalar-tensor theory of gravity with $w=\text{const.}$ and $V=0$, while the Bianchi cosmological models are anisotropic, which is suitable for showcasing the anisotropic local scaling transformations.

The connection dynamics for  Brans-Dicke gravity in the Bianchi I spacetime can be found in \cite{Bianchi}. Basically, the Bianchi I  model is a class of cosmological models with three different scale factors $a_i,~(i=1,2,3)$ along three spatial directions $x_I$, and metric for the Bianchi I model reads
\begin{equation}
ds^2=-N^2dt^2+a_1^2dx_1^2+a_2^2dx_2^2+a_3^2dx_3^2,
\end{equation}
where $(t,x_i)$ are Cartesian coordinates on the spacetime manifold and $a_1(t),a_2(t),a_3(t)$ are the directional scale factors. Generally speaking, in the Bianchi cosmological models, the spatial homogeneous hypersurfaces have the three-dimensional group of isometry which corresponds to the Bianchi classification of three-dimensional Lie algebras. Due to  these isometries, one can fix a fiducial spatial metric  so as to trivialize the Gaussian and vector constraints. More explicitly, one can fix a fiducial metric for the flat spatial hypersurface of the Bianchi I model as
$
ds^2_0=dx_1^2+dx_2^2+dx_3^2,
$
which determines the fiducial spatial triad $\mathring{e}_i^a$ such that $E^a_i=a_{(i)}\mathring{e}_i^a$ and $[\mathring{e}_i^a,\mathring{e}_j^a]=0$. Then, define the symmetry-reduced Ashtekar connections and densitized triads as
\begin{equation}\label{43}
A_a^i=c_iV_0^{-1/3}\mathring{e}_a^i,\quad \tilde{E}^a_i=p_iV_0^{-2/3}\sqrt{\mathring{q}}\mathring{e}_i^a,
\end{equation}
where $V_0=\ell_0^3$ with $\ell_0$ the length of a fiducial cell (which is assumed to be equal along the three directions). With the definition \eqref{43}, the canonical pair $(A_a^i,\tilde{E}^a_i)$ can be effectively changed to the canonical pair $(c_i,p_i)$.  Now that the Bainchi I model is spatially flat, we have vanishing spatial connection, $\Gamma^i_a=0$, and as a consequence of \eqref{66}, we see that  $\tilde{K}_a^i=\frac{1}{\gamma}A_a^i=c_i\mathring{e}_a^i/(\ell_0\gamma)$.

Now we only need to consider the Hamiltonian constraint. Since Brans-Dicke gravity is a special case of scalar-tensor gravity, the Hamiltonian constraint is still \eqref{10}, but with $V=0$. If we still assume the local scaling transformations \eqref{17} and \eqref{18}, then in the degenerate limit $\delta\rightarrow0$, we get the simplified constraint \eqref{23}, which can be rewritten in terms of $(c_i,p_i)$ as
\begin{equation}\label{4343}
\mathcal{C}'=-\frac{\mathring{q}}{2\phi\gamma^2\ell_0^6}(c_1p_1c_2p_2+c_2p_2c_3p_3+c_1p_1c_3p_3)+\frac{w+\frac{3}{2}}{2\phi\underline{N'}^2}(\partial_0\phi)^2.
\end{equation}
After smearing with $\underline{N'}$, we can readily obtain the Hamilton equations of motion for $c_i,p_i,\pi$. It is easy to see that the Hamilton equation of motion for $\phi$ vanishes as there is no $\pi$ in  \eqref{4343}, and hence $\phi$ is a constant of motion, which is the opposite to the case of Bianchi I spacetime in GR minimally coupled to a scalar field (cf. \cite{WE18}). Note that the first term in \eqref{4343} only differs from the GR case \cite{AWE09} by a factor of $\frac{1}{\phi}$.

Let us go back to the local scaling transformations \eqref{17} and \eqref{18}. Comparing these transformations to  \eqref{43}, we can see that the  scaling transformations should be on $c_i$ and $p_i$, because other factors in \eqref{43} are fixed by the fiducial metric. However, the indices in $c_i$ and $p_i$ are the internal indices matching the directions of the fiducial metric, and there is no spacetime index $a$ in them. Since $E^a_i=a_{(i)}\mathring{e}_i^a$, the local scaling transformation $E^i_a=\delta_{(a)} {E'}^i_a$ actually transforms the fiducial triad $\mathring{e}_i^a=\delta_{(a)}\mathring{e}_i^a$, but we can take the $\delta_{(a)}$ as a scaling transformation  $\delta_{(i)}$ for the directional scaling factor $a_i$, i.e. $a_i=\delta_{(i)}a'_i$, as long as the spatial direction $a$ matches the fiducial direction $i$.
Recall that the variable $p_i$ is defined by the directional scale factors \cite{AWE09}, for example, $p_1=\text{sgn}(a_1)\abs{a_2a_3}\ell^2$ and likewise for the other two, where $\text{sgn}(a_1)=1$ if $E^a_1$ is parallel to $\mathring{e}^a_1$ and $-1$ if they are antiparallel. Now considering the local scaling transformations] on the directional scale factors, $a_i=\delta_{(i)}a'_i$, we see that 
\begin{equation}p_i=\frac{\delta}{\delta_{(i)}}p'_i.\end{equation} 
Therefore, in order to conform to the local scaling transformations \eqref{17} and \eqref{18}, we require the matching of spatial and fiducial directions. On the other hand, the variable $c_i$ is given by the time derivative of the directional scale factor \cite{AWE09}, $c_i=\frac{\gamma}{\ell^2(a_1a_2a_3)}\frac{da_i}{dt}$. Again, by using the scale transformation  $a_i=\delta_{(i)}a'_i$, we have 
\begin{equation}c_i=\frac{\delta_{(i)}}{\delta}c'_i.\end{equation} 
In this sense, the scale transformations of $c_i$ and $p_i$ are consistent with \eqref{17} and \eqref{18}, thereby justifying the above degenerate limit.
\section{Conclusion} \label{S5}
We have studied the degenerate limits of scalar-tensor theories of gravity in their connection dynamics formulation. We considered two distinct sets of degenerate limits. In the first limit, the resulting constraints for degenerate scalar-tensor gravity are greatly simplified and share formal similarities with the constraints of electric Carroll gravity. Consequently, the subsequent loop quantization of this degenerate scalar-tensor gravity becomes more tractable than that of the full theory. In the second limit, the resulting scalar constraint differs drastically from that of magnetic Carroll gravity, with the scalar field contributions mixing nontrivially with the geometric ones. Given the complicated structure of the Carrollian limits of $f(R)$ gravity \cite{TK24}, this complication is not unexpected and warrants further investigation.

In this work, we assumed that the scalar field does not change under the scaling transformations of the geometry. From the perspective of background-independent quantum gravity, it would be more intriguing to consider scalar fields that do transform under the scaling. If all fields in the scalar-tensor theory scale accordingly, it might be possible to relate the theory to so-called scale-dependent gravity (cf. the recent work \cite{SS}).

The analysis of degenerate limits of canonical gravity in terms of connection dynamics could be further generalized to other modified theories of gravity, such as higher-dimensional scalar-tensor gravity \cite{HMZ14}  and the vector-tensor gravity \cite{LM25}.
\section*{Acknowledgements} 
X.G. is supported by Yancheng Institute of Technology (xjr2024030). BB acknowledges the support from the Scientific and Technological Research Council of T\"urkiye (T\"UB\'ITAK) 3501 under project no. 125F279, the National Science Foundation of China (NSFC) by Grants No. 11875006 and No. 11961131013.


\bibliographystyle{apsrev4-2}

\end{CJK*}
\end{document}